\documentclass[structabstract]{aa}

\usepackage{graphicx}
\usepackage{txfonts}
\usepackage{natbib}
\usepackage{url}
\usepackage{numprint}		
\usepackage[squaren,cdot]{SIunits}	

\newcommand{\abs}[1]{{\vert} {#1} {\vert} }

\bibpunct{(}{)}{;}{a}{}{,} 

\begin{document}
   \title{Extreme coronagraphy with an adaptive hologram}
   \subtitle{Simulations of  exo-planet imaging}
   \author{D. Ricci
          \inst{1}
          \fnmsep
          \thanks{Boursier FRIA}
        \and
          H. Le Coroller
          \inst{2}
        \and
          A. Labeyrie
          \inst{3}
          }
   \institute{
   		D\'epartement d'Astrophysique, G\'eophysique et Oc\'eanographie, B\^at. B5C, Sart Tilman, \\
   		Universit\'e de  Li\`ege, \\
   		B-4000 LIEGE 1, Belgium \\
              \email{ricci@astro.ulg.ac.be}
         \and
		Observatoire de Haute Provence,\\
		F-04870 Saint Michel l'Observatoire (France) \\
		\email{herve.lecoroller@oamp.fr}
         \and
		Coll\`ege de France,\\
		11, place Marcelin Berthelot \\
		75231 Paris Cedex 05 \\
		\email{antoine.labeyrie@obs-azur.fr}
             }

  \abstract
  {}
   {We present a solution to improve the performance of coronagraphs for the detection of exo-planets.}
   {We simulate numerically several kinds of coronagraphic systems, with the aim of evaluating the gain obtained with an adaptive hologram.}
   {The detection limit in flux ratio between a star and a planet ($F_s/F_p$) observed with an apodized Lyot coronagraph characterized by wavefront bumpiness imperfections of $\lambda/20$ (resp. $\lambda/100$) turns out to be increased by a factor of $10^{3.4}$ (resp. $10^{5.1}$) when equipped with a hologram.}
   {This technique could provide  direct imaging of an exo-Earth at a distance of $11$ parsec with a $6.5\meter$ space telescope   such as the JWST with the optical quality of the HST.      }

   \maketitle


\section{Introduction}

Most of the 300 or so exo-planets discovered since 1995  \citep{quelo} have been detected by the radial velocity or the photometric transit methods. In a few cases, transits could be observed spectroscopically  and provided data on chemical composition and temperature \citep{swain}.  A few sufficiently bright planets were imaged from the ground, with adaptive optics in the near-infrared, first by \cite{chauvin}, and more recently by \cite{marois}, who removed most of the unwanted diffracted starlight by ``angular differential imaging''. Another such detection  \citep{kalas} was made, in yellow and red light, with the Hubble Space Telescope.
The direct imaging of weaker extra-solar planets,  including exo-Earths about $ 10^{10}$ times fainter than their parent star at visible wavelengths,  remains an enormous challenge since the residual starlight in the image must be removed very efficiently to allow the detection of faint  planet images.

\noindent  The coronagraphic techniques developed since the solar coronagraph of \cite{lyot}, particularly in recent years, have  significantly improved its nulling gain,   which was limited to about \numprint{10000}. The ``Lyot coronagraph'' was improved  by replacing the opaque mask with phase masks \citep{rodier, rouand}, which attenuated the background level below  $10^{-7}$ \citep{riaud}. The stronger chromatic dependance of phase masks could be mitigated with devices such as the achromatic annular groove phase mask \citep{mawet}.

\noindent One can also improve the existing coronagraphs and erase the rings of the stellar point-spread function (PSF) by apodizing the pupil, for example using prolate functions \citep{aime, soummer}. Unfortunately, a large fraction of the light is absorbed by the apodization mask. The loss is avoided with an apodizing device \citep{guyon, guyonangel} using a pair of distorted mirrors  to modify the light distribution across the pupil. With this loss-less achromatic apodization, exo-planets as faint as $10^{-10}$ can in principle be directly imaged.

\noindent   However, in practice, even  a theoretically perfect coronagraph is greatly affected by the residual bumpiness of the incoming stellar wavefront, caused by imperfect mirror polishing or residual atmospheric turbulence, which cannot be perfectly corrected by adaptive optics.   The residual halo of starlight in the image  typically exhibits a speckle pattern at a relative level much higher than $10^{-10}$.  Mar\'echal's  formula \citep{livrantoine}  shows that to detect an exo-planet $10^9$ times weaker than its parent star with a  $10\meter$ mirror operating at $1\micro\meter$, the RMS wavefront error among  $10\centi\meter$ patches, such as obtained by \numprint{10000} actuators, should remain below $0.5\nano\meter$.

\noindent Since such accuracy is not achievable with the present figuring techniques, one method to remove the residues is to add stages of adaptive elements. Several adaptive devices proposed in the literature \citep{codona, labeyrie, putnam} use the following interesting property: if the star is unresolved, the light of the speckle halo is coherent with that of the central source, usually absorbed by the Lyot mask. This light in principle can be made to interfere destructively with the speckle halo. It improves the detectability of the faint non-coherent exo-planet, thus relaxing the accuracy of the AO actuators.

Here, we present simulations with the adaptive hologram method proposed by \cite{labeyrie}. Located close to the Lyot stop, a hologram removes most of the residual star light by adding to it a phase-shifted copy of its wavefront. In Sect.~\ref{design} we describe the optical design of the holographic technique adapted to the traditional Lyot coronagraphs.
In Sect.~\ref{nuovo} we discuss the adaptive hologram technology. In Sect.~\ref{opti} we give an analytic description of the diffracted orders generated by the hologram and  discuss how to optimize the parameters of the system.  In Sect.~\ref{simulation} we show the results of our numerical simulations, discussing  the gain obtained  under different conditions: ideal ones (Sect.~\ref{ideal}) and with static wavefront bumpiness and photon noise (Sect.~\ref{mirror},~\ref{noise}). Finally, Sect.~\ref{acro} deals with the degrading effect of a star which is partially resolved, its possible mitigation,  and  the design of an achromatized version.
 \begin{figure}[t]
   \centering
   \includegraphics[width=8.8cm]{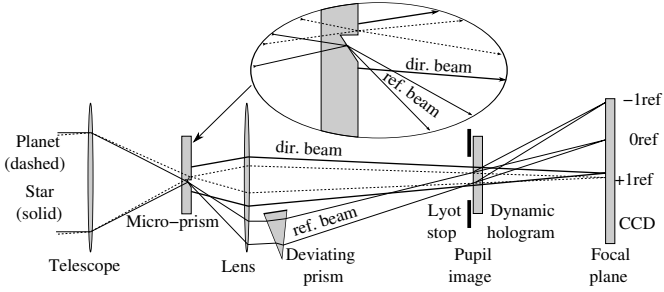}
      \caption{Holographic coronagraph using a modified Lyot train. The focal occultor is replaced by a deviating micro-prism which separates the star's  geometric beam for use as the holographic reference beam,  while preserving the outer diffracted rings and speckles, together with the images of planets, forming the direct beam. 
The micro-prism size, amounting in our simulations to $(8\lambda f/D)$ or $13.5\micro\meter$ in visible light at $F/3$, can vary between that of the Airy peak and a dozen rings. Equivalently, the micro-prism can also be replaced by a reflective focal plane with a hole selecting the reference beam.
       A larger deviating prism  deflects the reference beam in such a way that it intersects the direct beam in the relayed pupil image where the Lyot stop  is located.   A dynamic hologram,  located at the same position, first records the interference fringes produced by both intersecting beams, and then reconstructs a copy of the star's direct  wavefront, without the planet's contribution.   The hologram is made to introduce a $\pi$ phase shift in the reconstructed beam, causing  it to interfere destructively with the ``live''  stellar wavefront  (direct beam) also transmitted through it by zero-order diffraction, thus nulling the star's speckles and improving the planet's contrast in the relayed sky image detected by a CCD camera. The lens relays the telescope aperture at the pupil  plane and its focal plane at the camera. In the final image, the reference beam is  diffracted by the hologram into several orders ($-1\rm ref$, $0\rm ref$, and $+1\rm ref$ in the figure).  The order ``$+1 \rm ref$'' is a reconstruction of the speckled wavefront, phase-shifted by $\pi$.}
         \label{fig:lyot2}
   \end{figure}
 \begin{figure}[t]
   \centering
   \includegraphics[width=8.8cm]{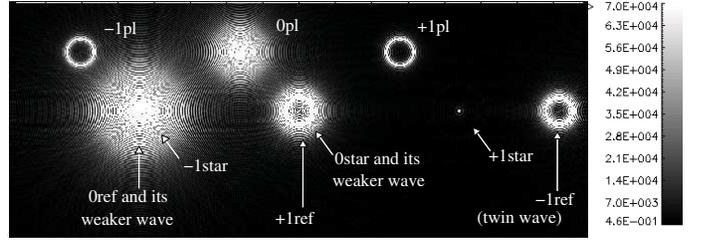}
      \caption{Simulation of an  imaged star-planet system in a Lyot coronagraph equipped with an adaptive hologram (see Fig.~\ref{fig:lyot2}). Here, for clarity, the planet is particularly bright and far from the parent star. The diffracted orders $-1$, $0$ and $+1$ of the planet are noted respectively $-1\rm pl$, $0\rm pl$ and $+1\rm pl$. The orders $-1$, $0$ and $+1$ of the star are noted respectively  $-1\rm star$, $0\rm star$ and $+1\rm star$. 
The order $+1\rm star$ looks faint in this figure because it is fainter than the direct flux of the star ( $0\rm star$) and the reference beam ($0\rm ref$).   The orders $-1$ (twin wave), $0$ and $+1$ of the reference beam are noted respectively  $-1\rm ref$, $0\rm ref$ and $+1\rm ref$. Star nulling is caused by the destructive interference  of its order $0$  (0star) and the order $+1$  of the reference beam (+1ref), both superposed on the  detector. Here, no $\pi$ phase shift is introduced in  the reference beam, so as to show the diffracted orders $+1\rm ref$, and $0\rm star$. The ``twin wave'' falls out of the detector at left , but here spuriously appears at right due to aliasing, itself caused by the moderate sampling which we had to use in the hologram. }
         \label{fig:ordini2}
    \end{figure}

\section{Coronagraph design improved with an adaptive hologram}
\label{design}
As previously described  \citep{labeyrielecoroller,labeyrie},  and shown in Fig.~\ref{fig:lyot2}, the focal occultor in a Lyot coronagraph can be built in the form of a micro-prism deflector preserving most star-light to produce an off-set reference wave.  It  interferes with the direct wave in order to generate  a hologram which is initially recorded  in the pupil plane near the Lyot stop, and then exploited for nulling most starlight while preserving the planet's image formation. Depending on the focal ratios, the reference beam may intersect the direct beam  at an angle  of several degrees.   We note $\psi_d$  and  $I_d = \abs{\psi_d}^2$  the complex amplitude of the direct beam and its intensity at a point $P$ of the hologram's plane (see Fig.~\ref{fig:lyot2}), while   $\psi_r$  and  $I_r = \abs{\psi_r}^2$ are the corresponding values for the reference beam.

\begin{figure}[t]
   \centering
   \includegraphics[width=6.3cm]{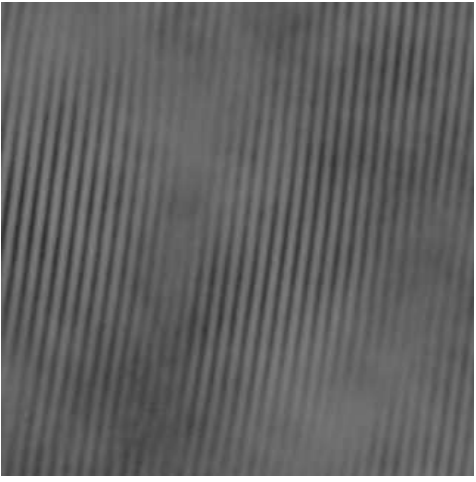}
      \caption{Small part of a hologram obtained with a laboratory simulator designed according to Fig~\ref{fig:lyot2}, suitably magnified to display the low-contrast fringes within the speckles. }
         \label{fig:verolo}
   \end{figure}

\begin{figure}[t]
 \centering
 \includegraphics[width=8.8cm]{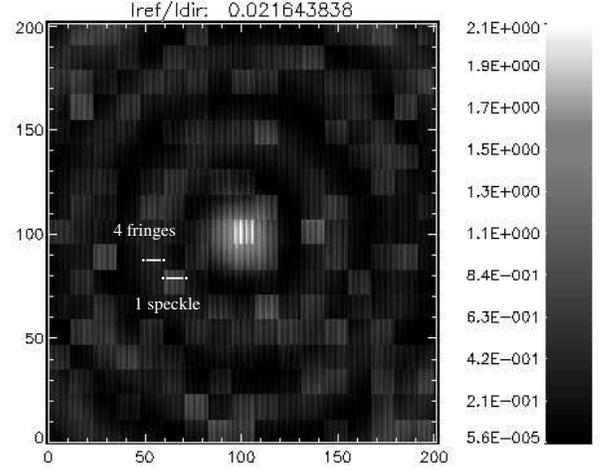}
    \caption{Numerical simulation of the central part of a hologram recorded  in the pupil with reference beam  at $65\degree$  incidence. For a contrasted display here, the  wave bumpiness is adjusted  at the $\lambda/40$ level, at $\lambda = 550\nano\meter$, and the reference beam intensity is much reduced,  to show both the  speckles and the finer fringes within them. There are approximately  four fringes per speckle, and four pixels per fringe period.  The total pixel count is  $1599\times 1599$, only   $200\times 200$ of which are displayed here.}
       \label{fig:holo}
 \end{figure}

\noindent The interference between the reference beam and the direct beam creates fringes in the speckles of the hologram (see~Figs.~\ref{fig:verolo} and~\ref{fig:holo}). The fringes of the recorded  hologram behave  like a grating, diffracting  several orders  which become focused in the focal plane (see Fig.~\ref{fig:ordini2}). We give a more complete and detailed  analysis of the creation and role of the diffraction orders  below. The complex amplitude resulting from the interference of the two beams at  point $P$   in the pupil is
\begin{equation}
	\psi 		= \psi_d + \psi_r
	,
	\label{eq:comp}
\end{equation}
while the intensity $I$ is
\begin {eqnarray}
	I 	&=& \psi\psi^\ast 	\nonumber \\
		&=& (\psi_d^{\phantom\ast} + \psi_r^{\phantom\ast})(\psi_d^\ast + \psi_r^\ast) 	 \nonumber \\
	  	&=& I_r + \psi_d^{\phantom\ast}\psi_r^\ast + \psi_d^\ast\psi_r^{\phantom\ast} + I_d	 \nonumber \\
	  	&=& I_r
		  	\left[ 1
		  	 +  \frac{\psi_r^\ast \psi_d}{I_r}
		  	 +  \frac{\psi_r\psi_d^\ast}{I_r}
		  	 +  \frac{I_d}{I_r}
		  	 \right]
			.
	\label{eq:int}
\end{eqnarray}

\noindent The hologram has an amplitude  transmittance $\tau = I^{\gamma / 2}$ where $\gamma$ is the classical intensity contrast  in photographic materials \citep{perez}. By illuminating the recorded hologram with the reference beam only, a reconstructed image of the star's speckles appears on the detector, noted $+1\rm ref$ on Fig.~\ref{fig:ordini2}. Using a reference beam phase-shifted by $\pi$, we obtain the same image phase-shifted by $\pi$ if the hologram is a positive print. Note that in the article of \cite{labeyrielecoroller} the $\pi$ shift was obtained with a negative hologram, which is equivalent.  Finally, the order 0 of the direct beam ($0\rm star$) adds destructively with the order ``$+1 \rm ref$'', thus nulling  the residual speckles of the star.

\noindent  The  planet's light, being incoherent with  the reference beam and little affected by the micro-prism,  is not  reconstructed by the hologram, and therefore escapes nulling. If we illuminate the recorded hologram with  the reference beam, now   phase-shifted by $\pi$, the transmitted  complex amplitude becomes:
\begin{eqnarray}
	\psi 	&=& \tau\left(\psi_d + \psi_r e^{i\pi}  \right)	
		 =  \tau\left(\psi_d - \psi_r           \right)		
		 .
	\label{eq:quattro}				
\end{eqnarray}
If $\gamma = 2$, the usual value considered optimal in holographic  practice, the product of  both factors having respectively four and two terms gives eight terms. As expected, since the reconstructed direct wave is $\pi$-shifted with respect to the transmitted direct wave, the corresponding pair of terms cancels.  Another pair of terms also cancels  for a related  reason. The expression thus simplified is:
\begin{equation}
	\psi 	= 	- I_r\psi_r + I_d\psi_d - \psi_r^2\psi_d^\ast + \psi_r^\ast\psi_d^2
	.
	\label{eq:ord}	
\end{equation}
The remaining  terms of Eq.~\ref{eq:ord} represent various orders diffracted by the hologram, locally behaving like a diffraction grating, and becoming separated on the  camera. Some of them  widely spread their speckled light  on the camera  (see Fig.~\ref{fig:ordini2}), thereby degrading the visibility of planet peaks, which are imaged through the hologram without being much affected by it.
It is of interest to discuss the impact of each term on the planet detection:
\begin{enumerate}
	\item 	the term $I_r\psi_r$ is  the transmitted  reference wave, producing an intense but sharp focus on the camera (order $0$ of the reference beam; 0ref in Fig.~\ref{fig:ordini2}). Apodizing the reference beam improves the nulling depth since its Airy  rings in its zero-order image pattern  become attenuated, and thus contaminate less the planet's image. Such apodization occurs naturally if the micro-prism separator selects the central Airy peak down to its  first dark ring. Additional spatial filtering can be achieved at the micro-prism if needed.
	\item 	the term $I_d\psi_d$ describes a wave that propagates close to the direct stellar residue wave ($I_r\psi_d$), nulled by the hologram, but with attenuated and spatially modulated intensity. This term is called the ``weaker wave'' in Fig.~\ref{fig:ordini2}.     The modulation causes some diffractive spreading on the camera, contaminating the planet image. This is mitigated by using a very intense reference beam, relative to the direct beam, so that $ I_d\psi_d \ll I_r\psi_d $. The intensity thus  achievable, as limited by the energy content of the Airy peak in the telescope's focal plane,  defines the maximal possible gain  with a hologram. This suggests that the maximum nulling achievable is: 

\begin{equation}
	G \approx \frac{\sum_{p} I_r^2}{\sum_{p} I_d^2}
	,
\label{gainmax}
\end{equation}
where the $\sum_{p}$ summing extends to all points of the pupil plane that are not blocked by the Lyot stop.
The numerical simulations support this estimate.
	\item	The term $\psi_r^\ast\psi_d^2$ is the order $+1$ of the direct beam ($+1\rm star$ in Fig.~\ref{fig:ordini2});
	\item	$\psi_r^2\psi_d^\ast$ is a wave similar to the direct beam (direct stellar residue wave), since  its complex amplitude is proportional to the conjugate of $\psi_d$. This last wave is termed the ``twin wave'' (order $-1$ of the reference beam; $-1\rm ref$ in Fig.~\ref{fig:ordini2}).  Its contamination is discussed in Sect.~\ref{opti}
\end{enumerate}
The order $-1$ of the direct beam ($-1\rm star$) cancels with the wave $I_d\psi_r$ that propagates close to the order $0$ of the reference beam.

\section{ Adaptive hologram technology }
\label{nuovo}
A dynamic, or adaptive, hologram works both as a wavefront sensor and an actuator array, thus behaving like the feedback loop of conventional adaptive optics. In the absence of rewritable holographic materials having enough light sensitivity and  response speed, the hologram's sensing and playing functions can also be achieved  by two separate components:   a camera  and  a Spatial Light Modulator (SLM).  To detect and process the light beam simultaneously, the camera can be fed by a beam splitter while the SLM is located in the relayed pupil, i.e. at the hologram position indicated in Fig.~\ref{fig:lyot2}. The latter can be driven by a video signal from the former, at the image rates of standard television, in the absence faster versions. Both components are small and commercially available devices which can fit, together with the beam-splitter, within a cubic inch.  Of interest are EM-CCD camera chips, incorporating electron multiplication which makes them nearly photon-limited at low light levels;  and SLMs such as those using a liquid crystal film on a silicon matrix, currently available with $1280\times 720$ pixels (see \url{www.cambridgecorrelators.com/products.html}).  Such existing components are readily usable to test a holographic coronagraph  on a laboratory bench and then on a telescope, particularly at red and  near infra-red wavelengths where the lifetime of speckles is longer. Inserting a computer or dedicated fast processor in the video connection is useful to adjust the $\pi$ phase shift, the gamma contrast, etc\dots\  of the printed hologram.   Much slower speeds suffice in space, where an artificial star such as a remote laser source can in principle be used to record the hologram.

\noindent With these existing components, a typical observing sequence  involves the near simultaneous  printing of the hologram with the live recorded pattern.   
\begin{figure}[t]
  \centering
  \includegraphics[width=8.8cm]{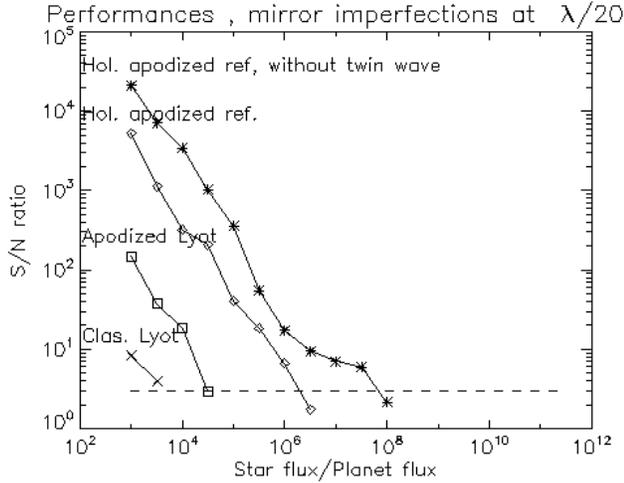}
     \caption{Approximate S/N ratio of planet detection  vs.  the star/planet flux ratio, in the presence of $\lambda / 20$ wavefront bumpiness at $\lambda = 550\nano\meter$ and without photon noise. (see Sect.~\ref{mirror}). 
     The detection limit (dashed) was chosen at S/N = $3$. We find a value of ${\sum I_r^2}/{\sum I_d^2} \approx 10^3$ for the solutions providing the hologram. 
     }
        \label{fig:lambda}
  \end{figure}


\section{Numerical simulations with optimized hologram parameters}
\label{opti}
 We adjusted  the size of the micro-prism to that of the fourth dark ring  in the Airy pattern; the planet was placed in the fifth  ring. This reproduces the angular separation between a star and an Earth-like planet located  at 1 UA from its parent star,  at $11\rm pc$ from us (like the Sun and Earth as seen from Gliese~436), and observed with a $6.5\meter$ telescope. In order to simplify the computations, we coarsely simulated the mirror bumpiness or turbulence  with a square grid of bumps with random amplitudes, and verified that the shape  of the phase cells on the mirror do not affect our results.

\noindent For minimal calculation noise, which represents possible problem given the high dynamic range considered, while keeping a reasonable computation time and usage of our 2GB RAM memory, we used  $1599\times1599$ pixel arrays,   with a pupil spanning $402$ sampling points.

\begin{figure}[t]
  \centering
  \includegraphics[width=8.8cm]{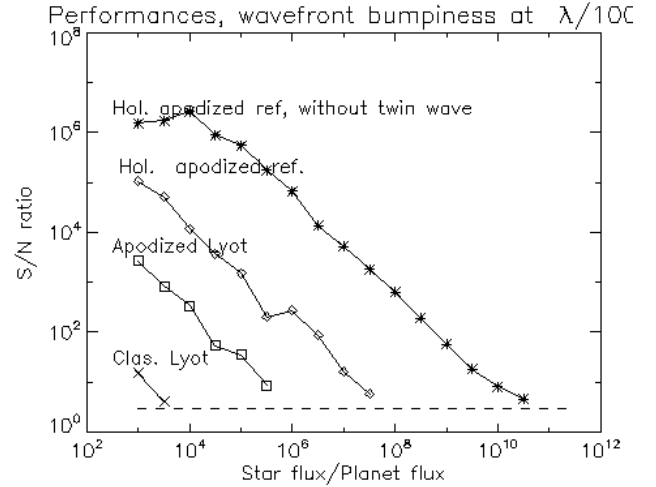}
     \caption{Approximate S/N ratio of planet detection  vs.  the star/planet flux ratio, in the presence of $\lambda / 100$ wavefront bumpiness, at $\lambda = 550\nano\meter$ and without photon noise (see Sect.~\ref{mirror}). 
     The detection limit (dashed) was chosen at S/N = $3$.We find a value of ${\sum I_r^2}/{\sum I_d^2} \approx 10^6$ for the solutions providing the hologram. 
     }
        \label{fig:lambda100}
  \end{figure}

\noindent The bumpiness of the incoming wavefront is converted into intensity speckles having random phases in the relayed pupil, i.e. in the hologram, in response  to the occultation of the Airy peak in the focal plane. In the pupil plane, the interference with the oblique reference beam generates  the finer fringes seen within each speckle (Figs.~\ref{fig:verolo} and \ref{fig:holo}). For a good sampling in the hologram, four pixels at least are needed per fringe period,  and simultaneously at least four fringes per speckle. Increasing the angle $\theta$ between the direct and reference beams increases  the number of fringes  per speckle, but decreases the number of pixels per fringe. It also increases the spacing of the diffracted orders in the focal plane, which improves the contrast of the planet's peak. In our simulation, we used about 4 pixels per fringe for an angle $\theta = 65\degree$ (see Figs.~\ref{fig:holo}). This corresponds to at least 4 fringes per speckle, if the typical scale of the bumpiness at the entrance pupil is larger than  $12\times 12$ pixels.
\noindent For optimal performance, it is important to adjust the following parameters:

\begin{itemize}
	\item In accordance  with the classical theory of holograms, the reference beam must be substantially more intense for a faithful wavefront reconstruction by the hologram, as apparent in the second  term  of Eq.~\ref{eq:ord}. In practice here, the  hologram's nulling gain is proportional to the square intensity ratio, i.e. $\approx10^{3}$ in our numerical simulation with wavefront bumpiness imperfections at $\lambda/20$ at $\lambda = 550\nano\meter$. This ratio is limited in practice by the energy content of the Airy peak, diverted by the micro-prism to form the reference  beam. The size of the micro-prism or attached filtering aperture, possibly apodized, should be optimized for maximal  light collection  while keeping a flat wave front. We found that an aperture sized like  the Airy peak is optimal in this respect and also to minimize the inner working angle.
	\item  As in conventional Lyot coronagraphs, the size of the Lyot stop was adjusted to  block the bright edge of the pupil image.
	\item Based upon Eq.~\ref{eq:ord}, we conclude that an optimal value for $\gamma$ is $2$ using a positive print, with a phase shifter then introduced in the reference beam during read-out, and $-2$ if it is negative.
	\item The twin wave is described by the term $\psi_r^2\psi_d^\ast$   in  Eq.~\ref{eq:ord}.  It gives a focal image identical to the direct wave, but shifted, and non-overlapping if the reference beam angle, with respect to the direct beam,  exceeds the apparent pupil size.  In our simulations, however, the pixel sampling of the hologram was insufficient to properly generate the twin wave, causing it to be aliased by the Fourier transform algorithm, and to appear in the final image at an incorrect  ``folded'' location which contaminated the planet image (See~Fig.~\ref{fig:ordini2}). The aliasing effect causes the twin wave to be nearer to the $0$ order  of the direct beam (and then to disturb the flux of the planet) than if it was not aliased.

In order to remove such numerical effects, we  simulated a suppressed twin wave by removing the corresponding term, in complex amplitude, in the numerical calculation, and found an improved detection sensitivity, part of which may result from the removed aliasing effect.

\noindent In this paper, we did not study the physical effect of the twin wave, but we know that the solution is intermediate between the case including the twin wave (C~in~Table.~\ref{tab:gain}) and the case where its effect is analytically subtracted (D~in~Table.~\ref{tab:gain}). More realistic simulations of the twin wave's effect are desirable in further work, and this may require non-FFT calculation methods \citep{soummeretal}.

\noindent Nevertheless, twin waves can be eliminated by using a thick hologram, also known as a Lippmann-Bragg hologram, in order to work in the case (D) of Table.~\ref{tab:gain}.
These have stratified nodal planes, rather than fringes, which selectively reflect a single first-order wave. Then, the hologram works like a ``blazed grating'', sending most light in the reconstructed image (order $+1\rm ref$).

\end{itemize}
\begin{table}
	\begin{center}
		\begin{tabular}{llllll}
\hline\hline
Conditions		& A) 		& B)		& C) 		& D) 		  	\\
\hline
Perfect conditions	& $10^{3.2}$	& $10^{11.0}$	& $10^{11.0}$	& $10^{11.0}$  	\\
$\lambda/100$		& $10^{3.5}$	& $10^{5.7}$	& $10^{7.6}$	& $10^{10.8}$   	\\
$\lambda/20$		& $10^{3.5}$	& $10^{4.5}$	& $10^{6.2}$	& $10^{7.9}$   	\\
\hline
\end{tabular}
\caption{Summary of the approximate limiting detection of the flux ratio $F_s / F_p$ with the different configurations mentioned in Sect.~\ref{simulation} without photon noise\setcounter{footnote}{1}\protect\footnotemark[\value{footnote}].}
		\label{tab:gain}
	\end{center}
\end{table}
\footnotetext{``$\lambda/100$'' and ``$\lambda/20$'' are referred to the gain after the introduction of wavefront bumpiness imperfection at $\lambda/100$ and $\lambda/20$ (Sect.~\ref{mirror}).}
\section{Gain evaluation}
\label{simulation}
Using the optimizations listed in Sect.~\ref{opti}, we were able to reproduce a wide range of coronagraph configurations, in order to compare their planet detection limit versus the "star/planet" flux ratio ($F_s / F_p$).

\begin{itemize}
	\item[(A)] Classical Lyot coronagraph;
	\item[(B)] Apodized Lyot coronagraph;
	\item[(C)] Apodized Lyot coronagraph with hologram and an apodized reference beam;
	\item[(D)] Apodized Lyot coronagraph  with hologram, apodized reference beam, and subtracted twin wave.
\end{itemize}

\noindent The performance is evaluated under three different conditions: ideal ones (no mirror imperfections and absence of photon noise); with the introduction of mirror imperfections and finally with photon noise in addition to the mirror imperfection.
These steps are treated in the following subsections, and summarized in Table~\ref{tab:gain} for the two first conditions.
The speckle noise was coarsely evaluated in the images using the following equation:
\begin{equation}
\frac{\bar I_{peak} - \bar I_{speckle} }{\sigma_{speckle}}	
\end{equation}
where $\bar I_{peak}$ is the mean intensity at the position of the planet; $\bar I_{speckle} $ is the mean intensity of the speckles near the planet position, and $\sigma_{speckle}$ is the corresponding root mean square fluctuation.

 \begin{figure}[t]
   \centering
   \includegraphics[width=8.8cm]{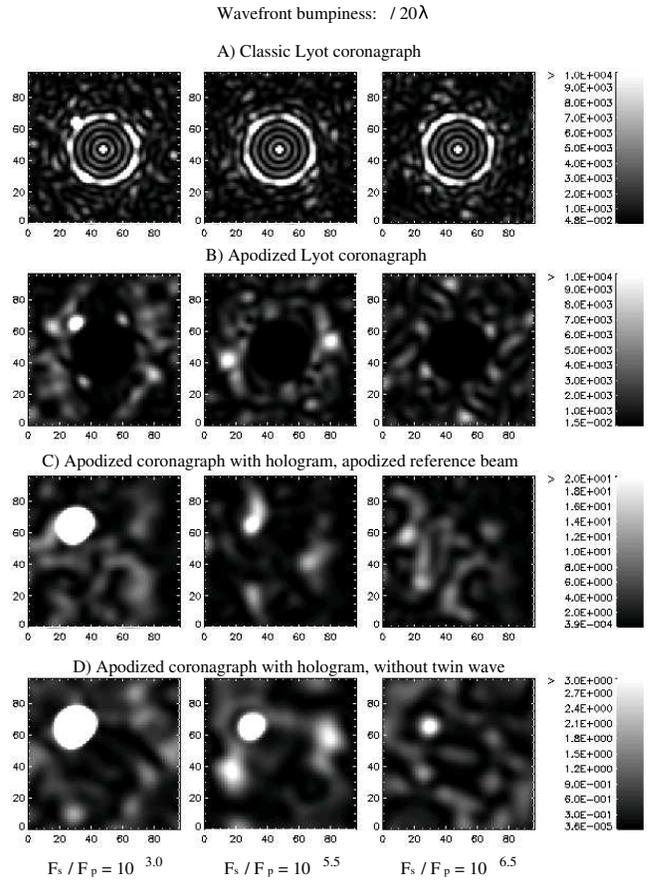}
      \caption{ The four  rows show simulated  images with the A, B, C and D coronagraph types mentioned in Sect.~\ref{simulation}. The star / planet flux ratio is $10^{3.0}$ in the left column, $10^{5.5}$ in the central column, and $10^{6.5}$ in the right column. The wavefront bumpiness is $\lambda / 20$ as described in  Sect.~\ref{mirror}. The intensity scales at right indicate the number of photons per pixel, for a star with magnitude $m_V=7$, a telescope diameter $D=6.5\meter$, $\lambda = 550\nano\meter$ and spectral bandwidth $\Delta\lambda = 10\nano\meter$. The exposure time is $60\second$ for both the hologram and the science camera }
         \label{fig:gain2}
   \end{figure}
\subsection{Gain with a perfect mirror}
\label{ideal}

The first results are obtained under  the assumption of perfect conditions, in the absence of mirror bumpiness and photon noise.

\noindent For the classical Lyot coronagraph (A), we obtain a detection limit in flux ratio of $F_s/F_p = 10^{3.2}$. In the case of the apodized Lyot coronagraph (B), the detection limit is increased to $10^{11.0}$. Introducing the hologram in the optical scheme with the apodization of the reference beam (C), the limit is $10^{11.0}$.  Our simulations appear to be limited to this range by numerical noise.  The  hologram becomes  most valuable  in the presence of mirror imperfections.

\subsection{Gain with bumpy mirror}
\label{mirror}

In order to test the performance under more realistic conditions, we have introduced a random static bumpiness on the incoming wavefront, with a  $\lambda/20$ peak-to-valley amplitude. We also performed simulations at $\lambda/100$, a situation intermediate between perfect conditions and $\lambda/20$ amplitude.  The results are shown in Figs.~\ref{fig:lambda}~to~\ref{fig:gain2}. 

\noindent The detection limit in flux ratio with a $\lambda/20$ wavefront bumpiness is  $F_s/F_p = 10^{3.5}$ with the classical Lyot coronagraph (A); $10^{4.5}$ with the apodized Lyot coronagraph (B); $10^{6.2}$ with the hologram and apodized reference beam (C); and  $10^{7.9}$ after subtracting the twin wave (D). At $\lambda/100$ we find $F_s/F_p = 10^{3.5}$ with the classical Lyot (A); $10^{5.7}$ with the apodized solution (B); $10^{7.6}$ with the hologram and apodized reference beam (C); and  $10^{10.8}$ after subtracting the twin wave (D).

\noindent The results with mirror bumpiness reveal the large  gain then obtained with the introduction of the hologram, especially when subtracting the twin wave. Indeed, with $\lambda/20$ wavefront bumpiness, it is not possible to detect a planet with an apodized Lyot coronagraph  if the ratio $F_s / F_p$ is larger than $10^{4.5}$ (see~Fig.~\ref{fig:gain2}). This limit is pushed to $10^{7.9}$ using the same coronagraph equipped with a hologram and after the subtraction of the twin wave. The detection limit is increased by a factor $10^{3.4}$. This number is about equal to the intensity ratio, predicted by the analytical estimation (see~Eq.~\ref{gainmax}).
  \begin{figure}[t]
    \centering
    \includegraphics[width=8.8cm]{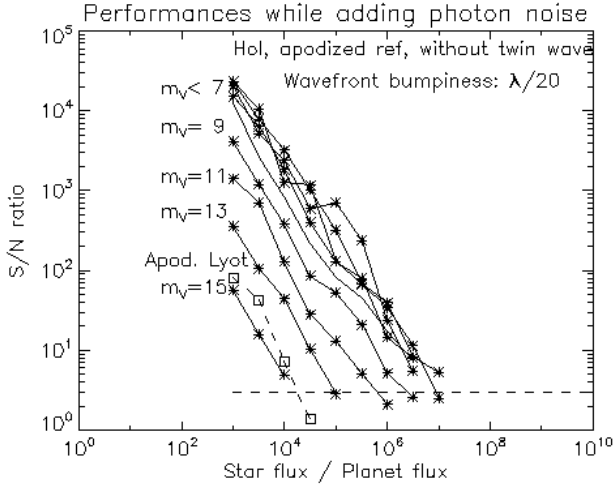}
       \caption{Approximate S/N ratio as a function of the star/planet flux ratio, evaluated with $\lambda / 20$ wavefront bumpiness and photon noise (see Sect.~\ref{noise}) for the coronagraph (D). curves are calculated for different stellar magnitudes, assuming $D=6.5\meter$, $60\second$  exposures, $\lambda = 550\nano\meter$ and  $\Delta\lambda = 10\nano\meter$. The performances of the apodized Lyot coronagraph at $m_V=7$ are also plotted (squares). The detection limit (dashed line) was chosen at S/N = $3$.}
          \label{fig:gain3}
    \end{figure}

\subsection{Gain with bumpy mirror and photon noise}
\label{noise}

After having tested the coronagraphs under perfect conditions and in the presence of mirror bumpiness, we added photon noise.  Unless the hologram is recorded with infinitely many photons, its recorded fringes are noisy, and this degrades the nulling depth in the camera image by creating a broad speckled halo, which degrades the visibility of the planet's peak.  It can be shown that the sensitivity limitation, regarding the detection of faint planets, is then ultimately the same as with a conventional adaptive system feeding a perfect coronagraph or apodizer, where the nulling depth is similarly limited by the number of photons detected by the wave sensor.  The calculation given  in the former case \citep{labeyrie} indeed also applies to the latter:  $N_p$ photons detected by either a hologram or a conventional wave sensor limit the achievable  peak/halo nulling depth $G$ to $G_{max} = N_p$. In both cases, as well as with a Mach-Zehnder interference nuller \citep{codona, putnam}, it follows  that recording and observing exposures lasting the same time leave  at least one star photon per speckle in the nulled camera image, which can be low enough to detect planet peaks containing several photons. The holographic nuller has no theoretical advantage in this respect.

\noindent Hybrid forms using both conventional adaptive optics for coarse wavefront correction, and a hologram for fine correction are also possible. They can use a single wave sensor such as a camera  located in the hologram  plane and serving as the holographic detector. Its signal can indeed provide a wavefront map to activate a deformable mirror in the entrance aperture. It can also be fed to a dynamic holographic plate, incorporating its own actuators, in the form of the fringe patterns acting as tiny gratings within each of its speckles. Like ordinary actuators in a servo loop, their performance is degraded by photon noise in the control signals, i.e. the recorded fringes with their ``frozen'' photon noise.   During the observation, after the hologram recording stage,  the more usual form of  ``live'' photon noise is also present in the recorded image and further contributes to degrading the detectability of a planet's peak if its level  is not much higher than the surrounding speckle peaks.

\noindent In our simulations, we have included both the ``frozen'' and the ``live'' photon noise contributions. We assumed that exposures of equal duration  served for the two phases of hologram recording and observation.  This may be optimal if the hologram is recorded with the same star which is subsequently observed.  If, however, a brighter reference star or a laser artificial star serves to record the hologram, the results given below may be considered as a lower limit since the photon noise of the hologram becomes attenuated in this latter case. Poisson-distributed photon noise was generated  in the hologram and the final image using a standard IDL routine, based  on C code \citep{numrec}.

  \begin{figure}[t]
    \centering
    \includegraphics[width=8.8cm]{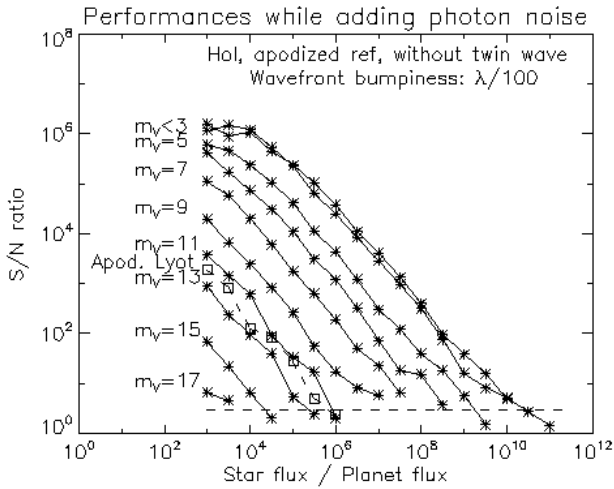}
       \caption{Approximate S/N ratio as a function of the star/planet flux ratio, evaluated with $\lambda / 100$ wavefront bumpiness and photon noise (see Sect.~\ref{noise}) for the coronagraph (D). curves are calculated for different stellar magnitudes, assuming $D=6.5\meter$, $60\second$  exposures, $\lambda = 550\nano\meter$ and $\Delta\lambda = 10\nano\meter$. The performances of the apodized Lyot coronagraph at $m_V=7$ are also plotted (squares). The detection limit (dashed line) was chosen at S/N = $3$.}
          \label{fig:gain4}
    \end{figure}


\noindent   Starting from the best solution, i.e. a hologram with an apodized reference beam and without a twin wave, we assumed $60\second$ exposures, both in recording the hologram and observing, using a $6.5\meter$ aperture in the V band ($550\nano\meter$), with a filter width of $10\nano\meter$. The results are shown in Figs.~\ref{fig:gain3}~and~\ref{fig:gain4} (simulations at $\lambda/20$ and $\lambda/100$ respectively). 

\noindent We see that the photon noise is negligible for stellar magnitudes brighter than 7 (at $\lambda/20$) and 3 (at $\lambda/100$); the hologram performance is then maximal, and independent from the star magnitude. In our simulation, those magnitudes correspond to about $\approx 3\times 10^6$ (at $\lambda/20$) and $\approx 3\times 10^8$ (at $\lambda/100$) photons per speckle in the hologram, and their performances are degraded below this value. 

\noindent Above magnitude 13 (at $\lambda/20$) and 11 (at $\lambda/100$), where the hologram has  $\approx 3\times 10^4$ (at $\lambda/20$) and $\approx 2\times 10^5$ (at $\lambda/100$) photons per speckle, the sensitivity gain vanishes with respect to a Lyot apodized coronagraph without a hologram.

\noindent  With a ground-based telescope, affected by ``seeing'',  bright stars may provide enough photons during brief exposures, shorter than the lifetime of ``seeing'', to activate both a conventional adaptive optics system giving a high Strehl ratio, and, within the coronagraphic attachment, the recording of a dynamic hologram. On fainter stars, a bright Laser Guide Star may similarly serve for both stages of adaptive correction.

\noindent In space, where the wavefront bumpiness is greatly reduced and varies much more slowly, some planet detection sensitivity can be gained by recording the hologram on a star brighter than the star observed, or on a ground-based laser source (which is not affected by turbulence if its emitting aperture  is unresolved).

It could be of interest if the hologram could attenuate the fixed stellar speckles escaping the adaptive correction. We attempted to address this by replacing  $\psi_d$ with a sum of a constant and a randomly variable term, in Eq.~\ref{eq:int}, but did not elucidate  the matter. Further work with simulations will be useful.

\section{Discussion}
\label{acro}

In the following subsections, two aspects for the future development of the instrument are considered: the chromatism of the coronagraph with a hologram;  and the problems introduced by a star that is poorly resolved.

\subsection{Achromatizing a hologram}
The fringe spacing in the hologram is normally proportional to wavelength but can  be made invariant if the deviating prism (See~Fig.~\ref{fig:lyot2}) in the reference beam is replaced by a diffraction grating. If the grating operates in the first order, its angular dispersion indeed increases the incidence angle of  light at increasing wavelengths.   This can reduce  the hologram's wavelength sensitivity, both during the recording stage and  the  observations, and increase the usable spectral bandwidth,  although the speckles contributed by the direct beam are also wavelength-dependant.  Further simulations, achieved  with a range of wavelengths,  would be of interest to specify the bandwidth then achievable.

Another approach involves a Lippmann-Bragg hologram, i.e. a thick hologram where the fringes are patterned as stratifications throughout the depth of the recording layer.  These can be wavelength-multiplexed, and can  simultaneously reconstruct the variously colored wavefronts that have been recorded.  Some recording materials, such as lithium niobate single crystals or polymeric compounds \citep{sishido}, are erasable and re-usable.  Their moderate recording sensitivity however may require a bright laser star.

\subsection{Resolved parent star}
The original Lyot coronagraph is highly tolerant of a star being resolved, as demonstrated by its initial success on the solar corona.  However, a  star which is slightly resolved by the telescope or hypertelescope \citep{labenovesei, coro} can degrade the hologram recorded with its light, and also degrade the nulling depth in the image cleaned by a pre-recorded hologram. In both cases, what matters is the invariance of the hologram, particularly in terms of fringe positions, with respect to a slight motion of a point star.
 Such motion produces identical translation shifts of the central and peripheral focal  patterns separated by the micro-prism.  The fringe pattern recorded in the hologram, located at a nearly infinite distance in the relayed pupil, is therefore nearly invariant. The pupil indeed remains fixed, and both interfering wavefronts reaching it become tilted by the same amount, while their phase difference is invariant. Their speckle detail is however slightly modified by the varying edge effects at the boundaries of the micro-prism and field lens. The hologram is thus expected to be somewhat tolerant of a resolved star serving to record it. Similarly, a hologram recorded  on a point source and then used to observe a resolved star is also expected to efficiently null its coronagraphic residue, since the live fringes remain contrasted. Further simulations and laboratory experiments will be of interest to estimate the tolerable apparent size of the star  vs. the desired nulling depth.


\section{Conclusions}
\label {conclusion}

Our diffractive analysis and simulations of a holographic coronagraph, and the comparison with other forms of adaptive coronagraphy,  show that the theoretical photon-limited sensitivity in detecting faint exo-planets is comparable. These methods will therefore need to be compared in terms of their technical implementations. Hybrid methods, for example combining an adaptive  mirror before the occultor and a hologram after, are likely to be similarly limited by the photon noise but may also have practical merits, such as relaxing the accuracy of the actuators \citep{putnam}.

\noindent  Depending on the holographic processes and materials that will become available, various forms of practical implementation may be of interest. With some re-writable holograms using amplifying photo-sensitive materials, no electronic image processing may be needed. Instead, and for a better flexibility, the recording and diffractive functions may be separated, using respectively a camera and a spatial light modulator.

With $\lambda / 20$ wavefront bumpiness,  the planet detection limit is improved, in terms of star/planet  flux ratio, from  $10^{4.5}$ to $10^{7.9}$.  In terms of photon noise, we found that the hologram improves the performance of the coronagraph if it is recorded with more than $\approx 3 \times 10^4$ photons per speckle. With $\lambda / 100$ wavefront bumpiness,  the detection limit improves from  $10^{5.7}$ to $10^{10.8}$, and the hologram improves the performance of the coronagraph if it is recorded with more than $\approx 2 \times 10^5$ photons per speckle.

\noindent We also discussed ways of making this holographic technique achromatic, and its tolerance for a poorly resolved star. In order to reach the best results, a thick Lippmann-Bragg hologram is of interest to remove the twin wave, but it restricts the range of available holographic components.
Following laboratory simulations, already initiated, tests on ground-based adaptive telescopes may be possible with available camera and Spatial Light Modulator components.
  \begin{acknowledgements}
  The research was supported by a collaboration between OHP -- Observatoire de Haute Provence and ARC -- Action de recherche concertée (Communauté Française de Belgique -- Académie Wallonie-Europe). We wish to thank the anonymous referee for careful review, stimulating remarks and suggestions. We also thank Jean Surdej who made  this collaboration possible,  Pierre Riaud for his helpful comments and Olivier Guyon for providing the apodization table.
  
  \end{acknowledgements}
  
\bibliography{biblio} 
\bibliographystyle{aa} 

\end{document}